\documentclass[%
reprint,
superscriptaddress,
 amsmath,amssymb,
 aps,
prb,
]{revtex4-2}

\usepackage{graphicx}% Include figure files
\usepackage{dcolumn}% Align table columns on decimal point
\usepackage{bm}% bold math
\usepackage{braket}
\usepackage{amsmath,mathtools}
\usepackage{siunitx}
\usepackage[section]{placeins}
\usepackage{comment}
\usepackage{mathrsfs}
\usepackage{rsfso}
\usepackage{lipsum}
\usepackage[T1]{fontenc}
\begin{document}

\preprint{APS/123-QED}

\title{Bosonic quantum error correction with microwave cavities for quantum repeaters}% Force line breaks with \\

\author{S.~Siddardha Chelluri}\email{schellur@uni-mainz.de}
\affiliation{Institute of Physics, Johannes-Gutenberg University of Mainz, Staudingerweg 7, 55128 Mainz, Germany}

\author{Sanchar Sharma}
\affiliation{Laboratoire de Physique de l’\'{E}cole Normale Sup\'{e}rieure, ENS, Universit\'{e} PSL, CNRS, Sorbonne Universit\'{e}, Universit\'{e} de Paris, F-75005 Paris, France}

\author{Frank Schmidt}
\affiliation{Institute of Physics, Johannes-Gutenberg University of Mainz, Staudingerweg 7, 55128 Mainz, Germany}

\author{Silvia Viola Kusminskiy}
\affiliation{Institute for Theoretical Solid State Physics, RWTH Aachen University, 52074 Aachen, Germany}
\affiliation{Max Planck Institute for the Science of Light, Staudtstraße 2, 91058 Erlangen, Germany}

\author{Peter van Loock}\email{loock@uni-mainz.de}
\affiliation{Institute of Physics, Johannes-Gutenberg University of Mainz, Staudingerweg 7, 55128 Mainz, Germany}

\date{\today}% It is always \today, today,

\begin{abstract}

Long-distance quantum communication necessitates the use of quantum repeaters, which typically include highly coherent quantum memories. We provide a theoretical analysis of the secret key rates for a quantum repeater system incorporating bosonic error correction and memory components. Specifically, we focus on the application of Binomial codes for two repeater segments. Using these codes, our investigation aims to suppress memory loss errors that commonly affect bosonic systems such as atomic gases, excitations of spin ensembles, and microwave cavities, in contrast to dephasing errors in single-spin memories. We further discuss a physical implementation of such a quantum repeater comprising a microwave cavity and a superconducting transmon, capable of state engineering with high fidelities ($>97\%$) and logical Bell state measurements for successful entanglement swapping. As an alternative approach, we also discuss a realization in the all-optical domain.

%Additionally, we introduce an algorithm designed to engineer highly non-Gaussian photonic states within microwave cavities, a crucial requirement for encoding quantum states. 
%Through simulations using realistic parameters, we achieve a fidelity exceeding 97\%. 
%Furthermore, we develop a new and efficient protocol for implementing Bell state measurements essential for entanglement swapping. 

% \begin{description}
% Secondary publications and information retrieval purposes.
% You may use the \texttt{description} environment to structure your abstract;
% use the optional argument of the \verb+\item+ command to give the category of each item. 
% \end{description}
\end{abstract}

\maketitle

%\tableofcontents

\section{\label{sec:level1}Introduction\protect\\}

The distribution of entanglement over long distances and between multiple parties, forming quantum networks, enables numerous applications such as the quantum Internet \cite{q_net}, uncompromised secure communication through the establishment of private keys \cite{bb84,e91}, distributed quantum computing \cite{dist_qc} and blind quantum computing \cite{blind}. However, the fiber channel transmittivity $\eta$ decreases exponentially with distance between parties, making long-distance quantum communication infeasible. This limitation can be addressed by the use of quantum repeaters \cite{rep} in which a channel is divided into multiple, say $n$, segments. With the addition of quantum memories, this division effectively increases the channel attenuation length by a factor of $n$, thus enhancing the transmittivity to $\eta' = \eta^{1/n}$.

% However, the fiber channel's transmittivity $\eta$ decreases exponentially with the distance between parties, making long-distance quantum communication infeasible. This limitation can be addressed by the use of quantum repeaters \cite{rep} in which a channel is divided into multiple, say $n$, segments. 
% %This division effectively reduces the total distance by a factor of $n$, thereby enhancing the transmission efficiency to $\eta' = \eta^{1/n}$ \cite{white}.
% This division implies that the entanglement needs to be maintained over a distance shorter by a factor of $n$, thereby enhancing the effective transmittivity $r\rightarrow r^{1/n}$ \cite{white}.

% The utilization of entangled quantum states for establishing a private key in communication channels achieves, in principle, a uncompromised security.

Quantum repeaters are typically classified into three generations based on their physical implementations \cite{rep_gen}: the first relies on memories and entanglement distillation, the second on memories and error correction, and the third exclusively on error correction.
As memories are crucial for the first two generations, various experimental platforms were explored as implementations, such as atomic ensembles \cite{dlcz}, color centers in diamonds \cite{diamond_exp, dia_rep}, and quantum dots \cite{white}. 
Due to decoherence in the memories, building quantum repeaters remains a significant challenge.
%Two main types of errors affect the memories used in repeaters: dephasing and loss, which depend on the experimental platform.
Depending on the platform, either dephasing or loss dominates the decoherence. In the latter case, since the primary imperfection in the entire repeater system is excitation loss — both in the optical channel and the memory node — it is crucial to investigate the impact of memory loss \cite{Lossy_PLOB}. While dephasing in memories has been extensively studied \cite{lars,memory_raz,memory_buffer,simon_rl,alena_2024,stav_sumeet,kenneth}, the loss of excitations in memories, particularly in the context of repeaters, has received relatively less attention, although prominent quantum repeater proposals would rely on memory systems subject to this type of error \cite{dlcz,sangouard_rmp}. To address loss errors, bosonic quantum error correction (QEC) can be applied, with examples including Binomial codes \cite{bin_code}, Cat codes \cite{cat_2013,cat_2014}, and Gottesman-Kitaev-Preskill (GKP) codes \cite{gkp}. Clearly, bosonic QEC is then also applicable to those repeater systems based upon bosonic memories \cite{stephan_gkp}. Bosonic qubit systems have also been applied in other areas, such as quantum simulation \cite{chen2024}.

%The performance of repeaters considering both memory loss errors and bosonic error correction has not been rigorously analyzed. Additional challenges include the lack of efficient methods for state preparation, syndrome detection, and entanglement swapping via logical Bell measurements in the context of bosonic QEC for specific physical systems. 

% Moreover, third-generation repeaters, though promising, are a distant prospect due to their complex optical state engineering and the requirement for closely spaced repeater stations.

% Repeater systems are evaluated by the rate at which secret bits of information are transmitted per second \cite{waiting}. 
The performance of a repeater system is typically evaluated in terms of the rate at which it can transmit secret bits of information per second \cite{waiting}. For third-generation repeaters, this figure of merit is relatively straightforward to calculate and has also been determined for repeaters using the Binomial code~\cite{Li2024}. Although third-generation repeaters are promising, they remain a fairly distant prospect due to the challenges with the complex optical state engineering and the need for closely spaced repeater stations~\cite{rep_gen}. In contrast, second-generation repeaters are much closer to practical hardware implementation and require significantly fewer stations to cover the same distance. However, calculating their transmission rates is far more challenging and becomes exponentially difficult, as the number of possible states in a repeater system grows exponentially with the number of repeater stations~\cite{waiting}.

In this work, we present a comprehensive analysis of a two-segment, second-generation repeater with loss errors where the information is encoded as a Binomial code in the memory. 
In these repeaters, photons remain unencoded during transmission, allowing for repeater stations to be spaced further apart than in typical third-generation systems.
Two error correction schemes are studied: one where error correction is applied once, and another where it is applied multiple times. Our figure of merit will be the secret key rate of quantum key distribution, which we numerically calculate for our repeater protocols. 

%A promising implementation of this quantum repeater is provided by a circuit quantum electrodynamics (cQED) setup~\cite{cQED_Rev1,cQED_Rev2}, consisting of a microwave cavity coupled to a superconducting transmon. cQED setups are interesting because of the long lifetimes of the cavity photons~\cite{HighCoh_MW}, required for hosting bosonic codes, and a nearly universal control of the cavity~\cite{UnivCont}, due to the highly non-linear nature of the accompanying transmon. This enables us to achieve the encoded state preparation and error correction syndrome measurement with cQED. Note that a similar approach has been considered for cavity QED (CQED) based on atom-light interactions \cite{Li2023}, with a focus on Cat codes in a third-generation quantum repeater. Here, we concentrate on cQED instead of CQED due to various bosonic QEC experiments in cQED, including Binomial~\cite{UnivBinom}, Cat~\cite{CatCode_Exp} and GKP~\cite{gkp_experiment} codes. Among the rotation-invariant bosonic codes, the orthogonality of the codewords in the Binomial code is beneficial \cite{Li2024}. Therefore, in this work, we focus on the Binomial code. 

A promising implementation of this quantum repeater is provided by a circuit or cavity quantum electrodynamics (cQED) setups~\cite{cQED_Rev1,cQED_Rev2}, consisting of a microwave circuit or a cavity coupled to a superconducting transmon. cQED setup are interesting because of the long lifetimes of the cavity photons~\cite{HighCoh_MW}, required for hosting bosonic codes, and a nearly universal control of the microwaves~\cite{UnivCont}, due to the highly non-linear nature of the accompanying transmon. We assume an ideal microwave-to-optical transducer, that can convert information from a cQED system to optical photons~\cite{QTrans_Rev,MWtoOpt_Lehnert,MWtoOpt_Exp}. Note that a similar approach has been explored for a cavity using atom-light interactions \cite{Li2023}, focusing on Cat codes in a third-generation quantum repeater, and in \cite{laha2025} for the engineering of Binomial code states. Here, we concentrate on transmon-enabled operations that have been used experimentally to demonstrate Binomial~\cite{UnivBinom}, Cat~\cite{CatCode_Exp} and GKP~\cite{gkp_experiment} codes.  
Among the rotation-invariant bosonic codes, the orthogonality of the codewords in the Binomial code is beneficial \cite{Li2024}. Another advantage is that  unlike other bosonic codes such as Cat or GKP codes, Binomial codes can be prepared without requiring an additional photon-number–related parameter (such as a phase-space displacement for Cat or a squeezing parameter for GKP), since they are truncated. Therefore, in this work, we focus on the Binomial code.

% [Here, you can write about optical cavity and also what do we like about it]
% Here we shall also briefly discuss an all-optical realization that, in particular, includes the linear-optics implementation of the entanglement swapping and error syndrome detection. 

The structure of this article is as follows: In Sec.~II, we concentrate on implementing QEC with Binomial codes for a second-generation quantum repeater protocol, and calculate the corresponding secret key rates. Section III introduces a deterministic algorithm for designing Binomial codeword states, syndrome detection, and logical Bell measurements for entanglement swapping in a microwave cavity system. Additionally, in Sec.~IV, we propose a protocol for logical-level entanglement swapping, achieving 50\% efficiency with linear optics. All numerical simulations and results are presented in Sec.~V.

% \subsubsection{physical systems and error models}

% The fundamental components to build a quantum repeater are quantum memories and qubits (blue and white circles in Fig. \ref{fig:bqec_two_segment}). There are several experimental platforms like NV centers, SiV centers, Qdots, Ions, Atoms etc to build quantum memories. However, in the subsequent sections, we will only discuss a repeater based on atomic ensembles. Photons, in which bit information is encoded within any two degrees of freedom, serve as the qubits as they are ideal for long distance transfer.

% \subsection{Bosonic quantum error correction}%

% As some of the experimental platforms such as atomic ensembles has loss as a dominant error (not dephasing). The storage time of quantum memories built of the spin ensembles is low and therefore the idea is to use Bosonic Quantum Error Correction (BQEC) to solve this problem. There are several BQEC codes but we choose here the single mode Binomial code \cite{bin_code} as we also show how to generate encoded states (state engineering) and perform syndrome measurement and entanglement swapping using linear optics.

%\section{\label{sec:level2}second generation quantum repeater loss model and QEC}
\section{\label{sec:qr_skr_theory}Quantum repeaters with QEC for lossy memories}

\begin{figure*}[htbp]
\includegraphics[width=\textwidth]{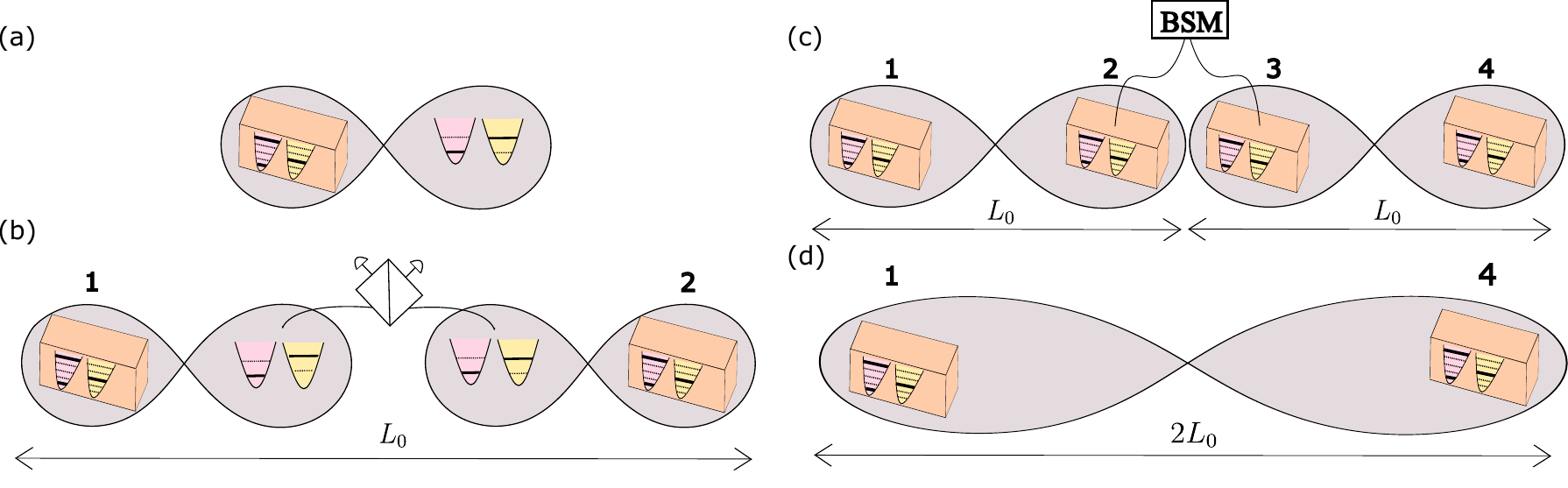}
\caption{\label{fig:bqec_two_segment} Two-segment, second-generation quantum repeater, with memory implemented as a microwave cavity (shown as an example). (a) Entanglement between a travelling photon and a microwave cavity occurs at all four memory units. (b) Entanglement within a single repeater segment of distance $L_0$ is established between two memory units via projection measurement (using beam splitters) on travelling photons, which are conducted independently in both repeater segments. (c) Entanglement swapping occurs locally between two memory units at the middle station of the two segments via a logical Bell state measurement (BSM). (d) The final entangled state is achieved between distant memory units 1 and 4. Throughout these steps, the superposition states of the photon are represented by yellow and pink photons, though only one entangled pair of photons is distributed per round of the protocol. Cuboids represent microwave cavities, and the encoded states of the photon are depicted by the thick lines in the harmonic oscillator.}
\end{figure*}

A quantum repeater involves dividing the channel between the communicating parties into small segments, distributing entanglement in those segments, and then using entanglement swapping to distribute the entanglement across all segments. For the rest of the paper, we consider a two-segment repeater, as illustrated in Fig.~\ref{fig:bqec_two_segment}, with a repeater node positioned in between (i.e., memory units 2 and 3 are located close to each other). The objective is to reliably distribute Bell pairs between memory units 1 and 4 (Fig.~\ref{fig:bqec_two_segment}). As physical memory units, we consider microwave cavities which are manipulated by superconducting transmons (not shown in the figure). Each segment requires two cavities at each end holding half of an entangled pair. Thus, there are two such cavities at each memory station, as shown in Fig.~\ref{fig:bqec_two_segment}(c).
In the first step, entanglement is independently distributed in the two segments, namely 1-2 and 3-4 (Fig.~\ref{fig:bqec_two_segment} (a),(b)). Since entanglement distribution is inherently probabilistic, the success probability for a single attempt is given by
\begin{equation}
    p = p_{\textit{link}} \, \eta_0,
    \label{eq:p}
\end{equation}
where \(p_{\text{link}}\) represents the intrinsic success probability of local hardware effects such as coupling efficiency, while the dominant factor \(\eta_0 = e^{-L_0/L_{att}}\) accounts for photon loss in the optical fiber channel. For example, at a distance of \(\SI{100}{\kilo\meter}\) and an attenuation length \(L_{\text{att}} = \SI{22}{\kilo\meter}\), the typical success probability is on the order of \(0.01\). Consequently, multiple rounds of transmission attempts are required to successfully establish entanglement across each segment. The time taken for each attempt is primarily given by the classical communication time $T_0= L_0/c$, where $L_0$ is the repeater segment length, and $c$ is the speed of light in optical fiber. Once a segment succeeds in generating entanglement, the corresponding qubits are stored in their respective memories. In case of failure, the memories are reinitialized and another attempt is made. This process continues until both segments achieve success. 

The second step is entanglement swapping between memory units 2 and 3 (Fig.~\ref{fig:bqec_two_segment}(c)). Entanglement swapping requires a joint Bell state measurement of one half from each of two separately entangled pairs (2 from 1-2 pair and 3 from 3-4 pair), resulting in entanglement between the remaining two halves in the respective pairs (1 and 4, Fig.~\ref{fig:bqec_two_segment}(d))\cite{es_swap,ekert}.

Since the success of entanglement generation in each segment is independent of the other segment, we define two geometrically distributed random variables: $n_{12}$ for segment 1-2 and $n_{34}$ for segment 3-4. These random variables represent the number of attempts required to distribute entanglement in each segment. As the generation of entanglement is probabilistic, let us assume that segment 1-2 establishes entanglement earlier than segment 3-4. As a consequence, memory unit 2 will be waiting to swap. In general, with either memory unit 2 or 3 waiting, the waiting time is defined as $n_{\rm diff} = |n_{12} - n_{34}|$ which follows a probability distribution $p_{\rm diff}$. During this waiting period, either memory units 1 and 2 or 3 and 4 experience dissipation, leading to an exponential decrease in the swapping probability \cite{simon}.
% \begin{equation}
%     P_s = \alpha_1 + \alpha_2 - \alpha_1\times \alpha_2
%     \label{def:Ps}
% \end{equation}
% where $\alpha_1 = \eta_m \times e^{-\frac{T_0}{\tau_{coh}}n_{diff}}, \alpha_2 = \eta_m \times e^{-\frac{T_0}{\tau_{coh}}}$ and $\eta_m$ represents the efficiency of a memory unit and $\tau_{coh}$ denotes the coherence time of the memory unit. 

To mitigate memory loss, we encode the memory qubit using a Binomial code (see Eqs.~(\ref{def:Low_Bin}), (\ref{def:High_Bin})) during the waiting period, thereby effectively enhancing the coherence time of the waiting memory units. The key performance metric for evaluating a quantum repeater is the (asymptotic) secret key rate, defined as
\begin{equation}
   S = r \times R.
   \label{def:SKR}
\end{equation}
Here, $R$ is the raw rate, i.e., the total number of bits transmitted per second. The fraction of the secret bits to the total bits transmitted is called secret key fraction and, considering the BB84 protocol, is given by
\begin{equation}
    r = 1-h(\langle e_x \rangle)-h(\langle e_z \rangle), \label{def:SKF}
\end{equation}
where the average is calculated by weighing it with the probability distribution function $p_{\rm diff}$ on condition that swapping is successful. Here $h(p)$ is the binary entropy function
\begin{equation}
    h(p) = - p\log p - (1-p) \log(1-p),
\end{equation}
and the quantum bit error rates (QBERs) $e_x$ and $e_z$ are defined as (assuming the desired state to be shared is $\ket{\psi^+}$)

\begin{equation}
   e_z  = \bra{00} \rho_{14}  \ket{00} + \bra{11} \rho_{14} \ket{11},
\end{equation}
and
\begin{equation}
    e_x  = \bra{+-}\rho_{14}\ket{+-} + \bra{-+}\rho_{14}\ket{-+}.
    \label{eq:ex}
\end{equation}
% \begin{equation}
%    e_z  = \bra{\overline{00}} \overline{\rho}_{14}  \ket{\overline{00}} + \bra{\overline{11}} \overline{\rho}_{14} \ket{\overline{11}},
% \end{equation}
% and
% \begin{equation}
%     e_x  = \bra{\overline{+-}}\overline{\rho}_{14}\ket{\overline{+-}} + \bra{\overline{-+}}\overline{\rho}_{14}\ket{\overline{-+}}.
%     \label{eq:ex}
% \end{equation}
where $ \rho_{14} $ is the density matrix of the state between memory units 1 and 4 (Fig.~\ref{fig:bqec_two_segment}).
To this end, we compare $S$ between cases with and without encoding to assess the impact of error correction on the memories.

% \begin{figure}[h]
% \includegraphics[trim=6.5cm 13cm 9cm 14cm]{figures/twosegements.pdf}% Here is how to import EPS art
% \caption{\label{fig:bqec_two_segment} Two-segment second generation quantum repeater. Blue dots indicate the memories and black lines indicate the transmission line.}
% \end{figure}

\subsection{Binomial code}

The Binomial code \cite{bin_code} is one of the bosonic codes to combat loss errors. The codewords of the lowest-order Binomial code (LBC), in the Fock basis, are
\begin{equation}
    \ket{\bar{0}}=\frac{\ket{0}+\ket{4}}{\sqrt{2}}, \;\;\; \ket{\bar{1}}=\ket{2}. \label{def:Low_Bin}
\end{equation}
This corrects single-excitation loss in the bosonic memories. The codewords of the next, higher-order Binomial code (HBC) are
\begin{equation}
    \ket{\bar{0}}_h=\frac{\ket{0}+\sqrt{3}\ket{6}}{{2}}, \;\;\; \ket{\bar{1}}_h=\frac{\sqrt{3}\ket{3}+\ket{9}}{{2}}. \label{def:High_Bin}
\end{equation}
This corrects for up to two excitation losses. We show the calculation of $S$ explicitly for the lowest order. We then numerically simulate both codes for a two-segment repeater and compare their secret key rates $S$.

\subsection{Secret key rate of a repeater with LBC}

In this section, we discuss the calculation of the secret key rate when the memory units are encoded with LBC. 

First, we generate one of the Bell states between the memory units 1-2 and 3-4. For any two memory units $\{i,j\}$, it is given by
\begin{equation} 
    \Ket{\psi^+}_{ij} = \frac{\Ket{0}_i \Ket{1}_j + \Ket{1}_i \Ket{0}_j}{\sqrt{2}}.
\end{equation}
The encoded Bell states are defined as $\ket{\overline{\psi^+}}$ by a similar definition as above with the replacements $\ket{0,1}\rightarrow \ket{\overline{0},\overline{1}}$. We also define the corresponding density matrix $\bar{\rho}_{ij} = \ket{\overline{\psi^+}}_{ij} \bra{\overline{\psi^+}}_{ij}$.

Without loss of generality, we assume that the Bell state was successfully generated in memory units 1-2 first. While waiting for memory units 3-4, memory units 1-2 undergo amplitude damping given by the set of Kraus operators~\cite{neil}, ${E_k}$ for all $k\in\mathbb{N}$,
\begin{equation}
   E_k = \sum_n \sqrt{\binom{n}{k}} \sqrt{(1-\gamma)^{n-k} \gamma^k} \ket{n-k}\bra{n}\,,
   \label{eq:kraus}
\end{equation}
where $n$ is the number of excitations, $k$ is the number of excitations lost, and $\gamma$ is the probability of losing an excitation stored in the memory unit.

The state of the memory units 1 and 2 after being subject to the amplitude damping channel is
\begin{equation}
    \overline{\rho}^{dec}_{12} = \sum_{k_1,k_2 = 0}^4 E_{k_1,1} E_{k_2,2} \; \overline{\rho}_{12} \; E^\dagger_{k_1,1} E^\dagger_{k_2,2}\,,
    \label{eq:rho_dec}
\end{equation}
where $E_{k,j}$ denotes $k$ losses in mode $j$.

To mitigate the effect of dissipation, error correction on the encoded qubits should be performed before entanglement swapping. Syndrome detection can be done by a parity measurement as discussed in Sec.~\ref{sec:microwave_cavities}. The recovery operation to be performed depends on the syndrome outcome. If there were no losses, the recovery operation to be performed, up to linear order of $\gamma$, is 
% \begin{equation}
% \begin{aligned}
%        \mathscr{R}_0 &= (\sin(\log({1-\gamma}))) (\ket{\Bar{0}} \bra{E^0_{\bar{0}}} - \ket{E^0_{\Bar{0}}} \bra{\Bar{0}}) \\  
%    & + (\cos(\log({1-\gamma}))) (\ket{\Bar{0}}\bra{\Bar{0}}  + \ket{E^0_{\Bar{0}}} \bra{E^0_{\Bar{0}}}) \\
%    & +\ket{\Bar{1}}\bra{\Bar{1}}
% \end{aligned}
% \end{equation}
\begin{equation}
\begin{aligned}
       \mathscr{R}_0 &= \ket{\overline{0}}\bra{\bar{0}^*}+\ket{\overline{1}}\bra{\bar{1}^*},
\end{aligned}
\label{eq:zero_recovery}
\end{equation}
where $\ket{\bar{0}^*} =  \ket{0}+(1-\gamma)^2\ket{4}/\sqrt{2(1-2\gamma)}$ and $\ket{\bar{1}^*} =  (1-\gamma)\ket{2}/\sqrt{(1-2\gamma)}$ . Note that $\mathscr{R}_0 = \text{id}$ when $\gamma = 0$. If a single loss was found, the recovery operation is
\begin{equation}
   \mathscr{R}_1 = \ket{\overline{0}}\bra{3}+\ket{\overline{1}}\bra{1}.
\end{equation}
In $\mathscr{R}_0$, the lack of complete orthogonality between error words and codewords is the cause of the significant disparity in expressions between $\mathscr{R}_0$ and $\mathscr{R}_1$.

% The next step is to perform entanglement swapping. It is performed by projecting the joint state of memory units 2 and 3 onto the Bell basis, for e.g, 
% \begin{equation}
%     \overline{\rho}_{14} = \frac{\prescript{}{23}{\bra{\overline{\psi^+}}} \overline{\rho}_{1234} \ket{\overline{\psi^+}}_{23}}{\text{Tr}\left[\prescript{}{23}{\bra{\overline{\psi^+}}} \overline{\rho}_{1234} \ket{\overline{\psi^+}}_{23}\right]},
%     \label{eq:rho14}
% \end{equation}
% similary for the other terms, where $\Bar{\rho}_{1234}= \mathscr{R}^{\otimes 2}(\Bar{\rho}^{dec}_{12}) \otimes \Bar{\rho}_{34}$ and $\mathscr{R}$ is $\mathscr{R}_0$ or $\mathscr{R}_1$ depending on the error syndrome information. 

The next step is to perform entanglement swapping, which is achieved by projecting the joint state of memory units 2 and 3 onto the Bell basis. For example,  

\begin{equation}
    \overline{\rho}_{14} = \frac{\prescript{}{23}{\bra{\overline{\psi^+}}} \overline{\rho}_{1234} \ket{\overline{\psi^+}}_{23}}{\text{Tr}\left[\prescript{}{23}{\bra{\overline{\psi^+}}} \overline{\rho}_{1234} \ket{\overline{\psi^+}}_{23}\right]},
    \label{eq:rho14}
\end{equation}  

with similar expressions for the other Bell states. Here, $\overline{\rho}_{1234}$ is given by $\mathscr{R}^{\otimes 2}(\Bar{\rho}^{dec}_{12}) \otimes \Bar{\rho}_{34}$ and $\mathscr{R}$ is $\mathscr{R}_0$ or $\mathscr{R}_1$ depending on the error syndrome information. 

The raw rate for the case of two segments with memory decay is given by \cite{simon}
\begin{equation}
    R = \frac{\langle P_s \rangle}{\langle n_{max} \rangle_{} T_0},
    \label{eq:raw_rate}
\end{equation}
where $n_{max}$ is the maximum of $n_{12}$ and $n_{34}$ and where $P_s$ is the swapping probability
corresponding to the sum of the traces of the conditional states of the Bell measurement (the denominator in Eq.~(\ref{eq:rho14}))
for all four Bell measurement results. The average of $P_s$ has to be taken, because the extent to which 
the state $\overline{\rho}_{1234}$ is subject to memory loss depends on the random memory storage (waiting) time
$n_{\rm diff}$. However, we will make a distinction between the cases where the loss error correction is only performed once at the end after storage and where it is done repeatedly for multiple times at a frequency given by the elementary time unit $T_0$.
Both cases are treated numerically by our simulation.

The figure of merit $S$ is calculated from $R$ and $r$ (defined in Eqs.~(\ref{eq:raw_rate}) and (\ref{def:SKF}), respectively). The secret key rates for different regimes of $\gamma$ are compared for the cases of encoded and unencoded memories in Sec.~\ref{sec:plots}.

\section{Quantum Repeater with Microwave cavities}
\label{sec:microwave_cavities}

In this section, we discuss a physical implementation of the quantum repeater, introduced in Sec.~\ref{sec:qr_skr_theory}, using cQED setups. 
As discussed in Fig.~\ref{fig:bqec_two_segment}, there are two microwave (MW) cavities, say $A_1$ and $A_2$, placed at every local station, each of which are loaded with a transmon qubit $Q_{1,2}$.
The first step is the generation of an entangled MW-qubit state, using local operations as discussed below.
Subsequently, the state of the transmon qubit is transferred to a photon which is sent to the middle station as shown in Fig.~\ref{fig:bqec_two_segment}. 
Remote entanglement swapping is performed via a joint Bell-state measurement (BSM) on $A_1$ and $A_2$ in the Binomial-encoded basis.
We discuss below a scheme for BSM using a third transmon, $Q_3$, coupled to both MW cavities.

\subsection{Hamiltonian}
Here, we discuss the relevant Hamiltonians for our system, whose implementations are well-known in the literature~\cite{cQED_Rev2}.
The Hamiltonian for each cavity is that of a bosonic mode $H_i^{a} = \hbar\omega_{a,i} a_i^{\dagger}a_i$ for $i\in\{1,2\}$, where $\omega_{a,i}$ and $a_i$ are, respectively, the resonance frequency and the annihilation operator of $A_i$.
For simplicity, we consider the resonant case $\omega_{a,1} = \omega_{a,2} = \omega_a$.

Each of the transmons are anharmonic oscillators, modelled by the Hamiltonian 
\begin{equation} 
    \frac{H_i^{qs}}{\hbar} = \omega_{q,i} q_i^{\dagger} q_i - \frac{\alpha_i}{2} q_i^{\dagger}q_i \left( q_i^{\dagger}q_i - 1 \right),
\end{equation}
for $i\in\{1,2,3\}$ where $\omega_{q,i}$, $\alpha_i$, and $q_i$ are the resonance frequency, anharmonicity, and the annihilation operator of $Q_i$. 
Furthermore, each transmon can be externally controlled via the dynamic Hamiltonian
\begin{equation} 
    \frac{H_i^{qd}(t)}{\hbar} = \Delta_i(t)q_i^{\dagger}q_i + \varepsilon_i(t)q_i^{\dagger} + \varepsilon_i^*(t)q_i,
\end{equation}
where $\Delta_i(t)$ is the externally controllable detuning and $\varepsilon_i(t)$ is the excitation amplitude corresponding to $Q_i$.
We assume that the cavities and transmons are so far detuned that the coupling is negligible for no external detuning, i.e. $\Delta_i = 0$.
Changing the detuning $\Delta_i$ can bring $Q_i$ in resonance with one or both of the MW cavities, allowing for entangling operations.

The individual MW-qubit couplings are in the form of a beam-splitter Hamiltonian $H_i^{aq} = \hbar g_{aq,i}(q^{\dagger}_ia_i + q_i a_i^{\dagger})$ for $i\in \{1,2\}$ with coupling coefficients $g_{aq,i}$. 
The third transmon used for BSM is coupled to both MW cavities with an effective dispersive coupling $\chi$, modelled with 
\begin{equation} 
    H^{\rm disp} = \hbar \chi(t) q_3^{\dagger}q_3 \left( a_1^{\dagger}a_1 + a_2^{\dagger}a_2 \right). \label{Q3:DispHam}
\end{equation}
The parameter $\chi$ can be tuned via the detuning $\Delta_3$ \cite{cQED_Rev1, cQED_Rev2}.

\subsection{Entanglement generation}
We want to create an entangled microwave and optical photon pair, as discussed in Fig.~\ref{fig:bqec_two_segment}.
In the first step, we create an entangled state in each MW-transmon subsystem. 
In this simulation, we can focus on only one MW-transmon pair ignoring the other, so we suppress the index, e.g. $a_i \rightarrow a$ and $q_i \rightarrow q$.
Dissipation is added in standard Lindblad forms~\cite{BP_OQS} of cavity loss, transmon loss, and transmon dephasing with corresponding Lindblad operators being $a$, $q$, and $q^{\dagger}q$ respectively.
We consider nominal values for a cavity with $\omega_a = 2\pi \times \SI{6}{\GHz}$, and lifetime of $\SI{50}{\micro\second}$.
We consider a transmon with $\omega_q = 2\pi\times \SI{5}{\GHz}$, $\alpha = 2\pi\times \SI{300}{\MHz}$, lifetime $T_1 = \SI{10}{\micro\second}$ and total dephasing time $T_2^* = \SI{5}{\micro\second}$.
We assume a coupling rate of $g = 2\pi\times \SI{25}{\MHz}$, which is sufficiently small to ensure that there is negligible MW-transmon coupling in equilibrium as $g \ll \omega_a - \omega_q$. 
The coupling is then induced by an input of $\Delta(t)$ that brings the MW and transmon in resonance.
Complementarily, excitations in the MW-transmon subsystem are generated by exciting the transmon with the pulse $\varepsilon(t)$, as written above.

We adapt the code given in Ref.~\cite{SancharCode}, to find $\Delta(t)$ and $\varepsilon(t)$ that leads to the entangled state 
\begin{equation}
    \Ket{\psi} = \ket{\bar{0}}\ket{e} + \ket{\bar{1}}\ket{g},
\end{equation}
where the MW state is encoded. 
Our simulations show that this state can be generated with $>97\%$ fidelity. 
The fidelity is primarily limited by cavity dissipation, and thus is significantly higher for cavities with higher lifetimes.
The resonance frequencies have only a weak effect on the fidelity.

The state of the transmon can be swapped into a travelling photon via a microwave-to-optical converter, creating an entangled state between microwaves and photons \cite{QTrans_Rev,MWtoOpt_Lehnert,MWtoOpt_Exp}. We consider ideal transduction between microwave and optical photons. The feasibility and implications of this assumption are discussed in the next section.

\subsection{Joint Bell-state measurement}
Here, we discuss how to perform a joint BSM on the two MW cavities, $A_1$ and $A_2$, using the transmon $Q_3$ coupled to both of them as discussed above.
In the following, we ignore the first two transmons as they are expected to stay in the ground state. 

A BSM results in a 2-bit output, where each of the four possibilities correspond to the four joint MW Bell states, namely 
\begin{equation} \begin{aligned}
    \ket{\overline{\Phi^{\pm}}} &= \frac{\ket{\bar{0}\bar{0}} \pm \ket{\bar{1}\bar{1}}}{\sqrt{2}} \\
    \ket{\overline{\Psi^{\pm}}} &= \frac{\ket{\bar{0}\bar{1}} \pm \ket{\bar{1}\bar{0}}}{\sqrt{2}}
\end{aligned} 
\label{BSs}
\end{equation}
The state $\Ket{\bar{i}\bar{j}}$ refers to the first cavity being in the $i$th logical state and the second cavity in the $j$th state.
If an error has occurred on any of the cavities, their states will instead be $\ket{\bar{0}^*}$ or $\ket{\bar{1}^*}$.

First, a parity check~\cite{MW_Parity} is performed on each of the cavities to find if an error has occurred. 
If no error has occurred, then a general state in the logical subspace is
\begin{equation} 
    \Ket{\psi}_{a_1a_2q_3} = \sum_{ij}\alpha_{ij}\Ket{\bar{i}\bar{j}}\Ket{g},
\end{equation}
where $i,j\in\{0,1\}$. 

To separate the $\Psi$ Bell states from the $\Phi$ Bell states, we employ the following steps. 
First, we apply a Hadamard gate to the qubit $Q_3$ by choosing $\varepsilon_3(t)=-i\varepsilon(t)$ (in the  rotating frame) satisfying
\begin{equation}
    \int dt \varepsilon(t) = \frac{\pi}{4}, \label{eq:Hadamard}
\end{equation}
to create an initial qubit state $\propto \ket{g} + \ket{e}$. 
Next, the dispersive coupling is turned on by reducing the detuning $\Delta_3$ such that $\chi(t) = \chi_0$ for a time period of $t=\pi/(2\chi_0)$, and zero outside the time window.
The effect of this coupling is to apply the entangling unitary on the joint system, 
\begin{equation}
    U = \exp\left[i\frac{\pi}{2}\left(a_{1}^{\dagger}a_{1}+a_{2}^{\dagger}a_{2}\right)\Ket{e}\Bra{e}\right].
\end{equation}
This unitary applies a phase on the qubit, depending on the total number of microwave photons in both cavities. 
The phase is periodic with photon number $4$. 
After the two operations, the joint state of the two cavities and $Q_3$ becomes 
\begin{equation} 
    \sum_{ij}\alpha_{ij}\Ket{\bar{i}\bar{j}}\frac{\Ket{g}+(-1)^{i+j}\Ket{e}}{\sqrt{2}}. \label{eq:post_ent} 
\end{equation}
Third, we apply another Hadamard gate on the transmon to get the state
\[ \Ket{g} \left( \alpha_{00}\Ket{\bar{0}\bar{0}} + \alpha_{11}\Ket{\bar{1}\bar{1}} \right) + \Ket{e} \left( \alpha_{01}\Ket{\bar{0}\bar{1}} + \alpha_{10}\Ket{\bar{1}\bar{0}} \right). \]
Now, we can measure the qubit to separate the $\Psi$-states from the $\Phi$-states. 
The two measurement outcomes can be written in the Bell basis as
\[ \Ket{g}\left(A_+\Ket{\bar{\Phi}^+} + A_-\Ket{\bar{\Phi}^-}\right),\ \Ket{e}\left( B_{+}\Ket{\bar{\Psi}^+} + B_{-}\Ket{\bar{\Psi}^-} \right) \]
for coefficients $A_{\pm}$ and $B_{\pm}$ that can be found in terms of $\alpha$-coefficients.

On the binomial code, universal gate operations were demonstrated \cite{UnivBinom}. 
So, we can apply a Hadamard gate on the encoded qubits in both microwave cavities to get the two possibilities depending on the outcome of the above measurement,
\[ \Ket{g}\left( A_{+} \Ket{\bar{\Phi}^+} + A_{-} \Ket{\bar{\Psi}^+}\right),\ \Ket{e}\left( B_{+} \Ket{\bar{\Phi}^-}-B_{-}\Ket{\bar{\Psi}^-}\right). \]
We can again apply the above three steps to separate $\Psi^{\pm}$ from $\Phi^{\pm}$, finishing the Bell state measurement.

If an error has occurred on the first cavity, Eq.~(\ref{eq:post_ent}) will be modified as
\begin{equation} 
    \sum_{ij}\alpha_{ij}\Ket{\bar{i}^*\bar{j}}\frac{\Ket{g} - i(-1)^{i+j} \Ket{e}}{\sqrt{2}}.
\end{equation}
Then, we can apply a Hadamard gate preceded by a phase gate to cancel off the factor $-i$. 
Note that both the gates can be applied in one step by choosing $\varepsilon_3(t) = -\varepsilon(t)$ with the condition in Eq.~(\ref{eq:Hadamard}). 
Then, the rest of the process follows in the same way.

Similarly, if an error occurred on both cavities, Eq.~(\ref{eq:post_ent}) will be modified as
\begin{equation} 
    \sum_{ij}\alpha_{ij}\Ket{\bar{i}^*\bar{j}^*}\frac{\Ket{g} - (-1)^{i+j} \Ket{e}}{\sqrt{2}},
\end{equation}
and the same procedure as in the no-error case can be applied.

\subsection{Experimental considerations}

The coupling of a superconducting transmon and a microwave cavity has been implemented in numerous experiments over decades~\cite{cQED_Rev1, cQED_Rev2}.
Binomial codes and their universal control has been demonstrated in microwave cavities~\cite{UnivBinom}.
For our simulations of the generation of entanglement between microwave and transmon systems, we use the Hamiltonian given in Eq.~\eqref{Q3:DispHam} to describe the coupling between two microwave cavities mediated by a transmon. The coupling parameters are taken from experimentally demonstrated values~\cite{MW_MW_Coup}. In such a system, care must be taken to minimize cross-talk between different transmons. To achieve this, the transmons are typically kept off-resonant with the cavities, effectively switching the couplings off. The couplings are activated only within specific time windows, and only one coupling is switched on at any given moment. This approach effectively suppresses cross-talk between transmons.
% However, as different transmons are brought into resonance only at specific times, we expect the cross-talk to have a negligible impact.

We have assumed an ideal microwave-to-optical converter in our analysis.
While such converters are not yet realizable, there has been a significant progress in achieving partial entanglement between microwaves and optics \cite{QTrans_Rev,MWtoOpt_Lehnert,MWtoOpt_Exp}. 
Since such a conversion is being coveted for several applications in quantum computing, we expect such converters to be available in the future.

If the microwave-to-optical conversion is not ideal and can instead be modeled as a probabilistic process—where, for instance, the conversion succeeds with a probability of \(99\%\) and fails with \(1\%\)—its effect can be directly incorporated into the secret key rate (SKR) calculation. In this case, the conversion efficiency modifies only the entanglement generation probability within a repeater segment (see Eq.~\ref{eq:p}). Specifically, the intrinsic link success probability \(p_{\text{link}}\) in Eq.~\ref{eq:p} can be replaced by an effective probability
\[
p'_{\text{link}} = p_{\text{cov}} \, p_{\text{link}},
\]
where \(p_{\text{cov}}\) denotes the microwave-to-optical photon conversion probability during the state preparation stage. The entanglement swapping stage remains unaffected since it does not involve frequency conversion. Therefore, no additional modifications are required in our current theoretical model for evaluating the SKR. However, if the microwave-to-optical conversion cannot be described as such a simple probabilistic success-or-failure process, integrating its effect into the SKR analysis becomes significantly more complex.

%\begin{enumerate}
%    \item Apply a Hadamard gate to the qubit to get
%    \[ \sum_{ij}\alpha_{ij}\Ket{i_{L}j_{L}}\frac{\Ket{g}+\Ket{e}}{\sqrt{2}}. \]
%    \item Wait for the dispersive interaction effectively giving the unitary
%    \[ U=\exp\left[i\frac{\pi}{2}\left(a_{1}^{\dagger}a_{1}+a_{2}^{\dagger}a_{2}\right)\Ket{e}\Bra{e}\right],\]
%    with $a_{\mu}$ being the annihilation operator of $\mu$th cavity. This gives the state
%    \[ \sum_{ij}\alpha_{ij}\Ket{i_{L}j_{L}}\frac{\Ket{g}+\left(-1\right)^{i+j}\Ket{e}}{\sqrt{2}}. \]
%    \item Apply another Hadamard gate to the qubit to get
%    \[ \Ket{g}\left(\alpha_{00}\Ket{0_{L}0_{L}}+\alpha_{11}\Ket{1_{L}1_{L}}\right)+\Ket{e}\left(\alpha_{01}\Ket{0_{L}1_{L}}+\alpha_{10}\Ket{1_{L}0_{L}}\right). \]
%    \item Measure the qubit to separate the $\Psi$-states from the $\Phi$-states. In what follows, it is better to write the state in Bell basis. The two measurement outcomes are
%    \[ \Ket{g}\left(A_{+}\Ket{\Phi_{+}}+A_{-}\Ket{\Phi_{-}}\right),\ \Ket{e}\left(B_{+}\Ket{\Psi_{+}}+B_{-}\Ket{\Psi_{-}}\right). \]
%    \item Apply a Hadamard gate on the cavity to get either of the two possibilities:
%    \[ \Ket{g}\left(A_{+}\Ket{\Phi_{+}}+A_{-}\Ket{\Psi_{+}}\right),\ \Ket{e}\left(B_{+}\Ket{\Phi_{+}}-B_{-}\Ket{\Psi_{-}}\right). \]
%    \item Again follow steps 1-3 to get the two possibilities:
%    \[ A_{+}\Ket{g}\Ket{\Phi_{+}}+A_{-}\Ket{e}\Ket{\Psi_{+}},\ B_{+}\Ket{e}\Ket{\Phi_{+}}-B_{-}\Ket{g}\Ket{\Psi_{-}}.\]
%    \item Measure the qubit again to finally distinguish all the cases.
%\end{enumerate}

\section{Quantum repeater with QEC in the optical domain}

Another feasible experimental platform for quantum repeaters is optics. In an all-optical approach, we could bypass the need for a microwave-to-optical conversion. However, in this case, we need to store phase-sensitive photonic states in an optical cavity system. To a certain extent, this has been experimentally demonstrated already, at least for phase-sensitive states up to one photonic excitation \cite{Pvl_phase_locking,Pvl_storage,Pvl_storage_2}. Another benefit of this all-optical approach would be that the local state processing at each repeater node can be done much faster than with light-matter interactions. The hardest element in an all-optical domain is the state preparation. There are a few experimental demonstrations of generating higher Fock states \cite{optical_fock_1,optical_fock_2} which are crucial to generate Binomial codeword states. Concerning the logical Bell measurements, we show that entanglement swapping can be achieved with linear optics at a 50\% success rate using photon-number-resolving detectors (PNRDs), provided that a recovery operation is performed prior to the swapping process. The corresponding derivation is presented in App.~A, including a figure that shows that the overall behavior of the SKR remains the same.

% The logical Bell measurement is divided into two steps. In the first step we show how the states $\ket{\overline{\psi^{\pm}}} \propto \Ket{\overline{01}} \pm \Ket{\overline{10}}$ are distinguished from $\ket{\overline{\phi^{\pm}}} \propto \Ket{\overline{00}} \pm \Ket{\overline{11}}$. In the second step, we distinguish $\ket{\overline{\psi^+}}$ and $\ket{\overline{\psi^-}}$. After straightforward calculation from Eqs.~(\ref{def:Low_Bin},\ref{phiout}), we find that the total number of photons detected after the beam splitter distinguishes the $\ket{\psi}$ states from the $\ket{\phi}$ states. 

% In the second step, we look at the click patterns for the  $\ket{\overline{\psi^+}}$ state and the $\ket{\overline{\psi^-}}$ state, considering all possibilities. These patterns are shown in App.~\ref{appendixa} for both the cases of no loss and one loss, using the beam splitter transformation. By examining the equations in App.~\ref{appendixa}, it becomes evident that the click patterns differ between the input states $\ket{\overline{\psi^+}}$ and $\ket{\overline{\psi^-}}$, enabling us to distinguish between them. This same reasoning can be extended to the higher binomial codes with up to two losses.

\section{Simulation of Secret Key Rates}
\label{sec:plots}

Here, we discuss the simulation and numerical evaluation of secret key rates for both cases, unencoded and encoded (LBC and HBC). In essence, we need to calculate Eq.~(\ref{def:SKR}). However, there are two possibilities in the calculation process: implementing error correction only once, referred to as single-time error correction (SEC), or implementing error correction multiple times at regular intervals during the waiting time, referred to as multiple-time error corrections (MEC). The motivation for MEC is to keep the memory states intact for longer while they are waiting. However, a typical problem is that the recovery operators are only approximate to first order, making it unclear whether repeated application would actually be beneficial. To address this, we implemented MEC under the assumption of no additional overheads, in order to determine whether it is feasible even in the idealized case. A detailed theoretical analysis of MEC is provided in App.~\ref{appendixc}. The procedure to simulate the secret key rates $S$ with SEC is described in Sec.~\ref{sec:qr_skr_theory}. Since it is SEC, $\mathscr{R}$ is applied only once.  To numerically calculate the averages in Eq.~(\ref{eq:raw_rate}), a cutoff time on the memories (apart from the exponential decay) was implemented. The swapping probability $P_s$ is given by expressions such as the denominator of Eq.~(\ref{eq:rho14}). For the SEC case, the cutoff was selected based on the time step corresponding to $\gamma=0.5$. In the case of MEC, $\gamma=0.1$ with $100$ as an upper limit for the maximum number of times the recovery can be performed is considered. The parameters $L_{att}= \SI{22}{\kilo\meter}$, $c = \SI{2e8}{\meter\per\second}$ were used throughout. The secret key rates in both cases are calculated under the assumption of $100\%$ efficiency in the BSM during swapping. Three different coherence times are considered for the simulations, including realistic values from state-of-the-art experiments~\cite{HighCoh_MW} and possible future values. All simulations were performed in Mathematica.  
  
% The comparison of SEC and MEC is shown in Fig.~\ref{fig:SKR_plot_KvsM}.
%Two approaches are used to obtain the numerical results for the two segment repeater. First method is apply the error correction only once (during swapping) and second approach is to do it multiple times (theoretical analysis in appendix \ref{appendixc}). For the first approach the comparison of secret key rates for un encoded scheme and the encoded scheme is shown in figures \ref{fig:SKR_plot_eta_1} and \ref{fig:SKR_plot_eta_095}. Furthermore, both approaches are compared and the results for the secret key fraction are showed in figure \ref{fig:SKR_plot_KvsM}. 

\begin{figure}
\includegraphics[width=\columnwidth]{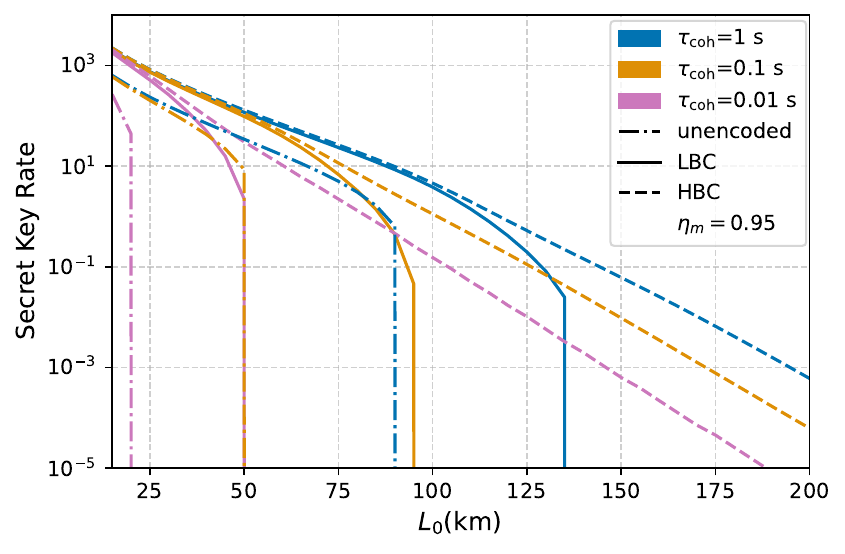}% Here is how to import EPS art
\caption{Comparison of secret key rates with different memory coherence times for unencoded, Lower Binomial Code (LBC) and Higher Binomial code (HBC) when SEC is performed. The memory interface parameter $\eta_m=0.95$.}
\label{fig:SKR_plot_eta_095}
\end{figure}

\begin{figure}
\includegraphics[width=\columnwidth]{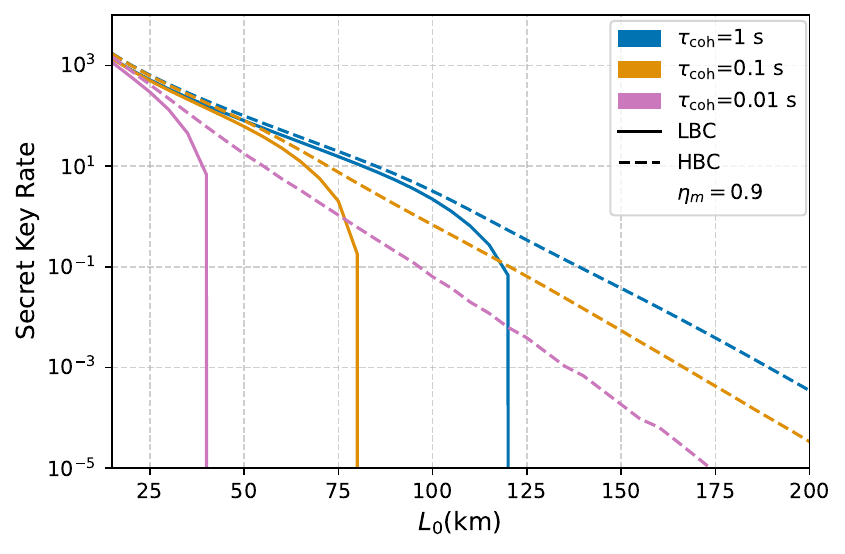}% Here is how to import EPS art
\caption{Comparison of secret key rates with different memory coherence times for Lower Binomial Code (LBC) and Higher Binomial code (HBC) when SEC is performed. The memory interface parameter $\eta_m=0.9$.}
\label{fig:SKR_plot_eta_090}
\end{figure}

In Fig.~\ref{fig:SKR_plot_eta_095}, we show the secret key rate as a function of the segment length, $L_0$ (range $10$-$200$ $\si{\kilo\meter}$), assuming a highly efficient memory interface, $\eta_m = 0.95$. We observe a clear improvement in the secret key rates of the encoded schemes in comparison with the unencoded one for all three different coherence times considered. For a more imperfect memory interface, $\eta_m = 0.9$, the results are shown in Fig.~\ref{fig:SKR_plot_eta_090}. They exhibit a similar trend but the unencoded case gives no positive secret key rates.  Interestingly, memories with error correction outperform the unencoded memories even with coherence times one order of magnitude higher. 

% Furthermore, only the HBC with a coherence time ($\tau_{coh}$) of $\SI{1}{\second}$ surpasses the scaling limit imposed by the maximum secret key capacity of the direct point-to-point link using optical fibers (PLOB bound) \cite{Plob}.

% \begin{figure}
% \includegraphics[width=\columnwidth]{figures/LOW_HIGH_UN_ENC_skr_tcoh_all_eta_1.pdf}% Here is how to import EPS art
% \caption{Comparison of secret key rates with different memory coherence times for unencoded, Lower Binomial Code (LBC) and Higher Binomial code (HBC). The memory interface parameter $\eta_m=1.0$.}
% \label{fig:SKR_plot_eta_1}
% \end{figure}

\begin{figure}
\includegraphics[width=\columnwidth]{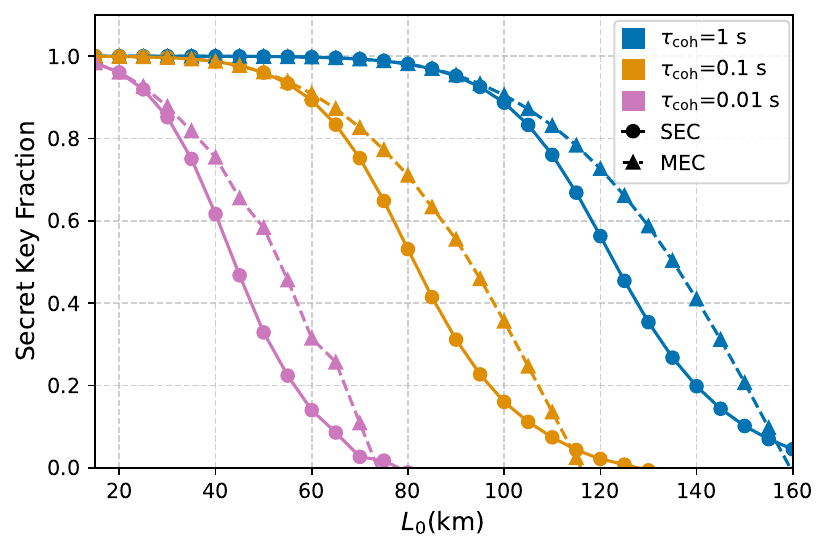}% Here is how to import EPS art
\caption{Comparison of the secret key fraction of LBC with different coherence times for the cases of SEC and MEC. The memory interface parameter $\eta_m=1.0$.}
\label{fig:SKR_plot_KvsM}
\end{figure}

In Fig.~\ref{fig:SKR_plot_KvsM}, we show the secret key fraction ($r$) as a function of the segment length $L_0$ for LBC showing the difference between the cases of SEC and MEC. As expected, allowing for multiple error corrections increases $r$ for long distances. For sufficiently short distances, a single recovery step is more appropriate. With increasing coherence times the distance at which SEC and MEC start to deviate is also increasing. Even though the value of $r$ is almost unity for the blue curves ($\tau_{coh}=\SI{1}{\second}$) at $\SI{100}{\kilo\meter}$ for both SEC and MEC, from Fig.~\ref{fig:SKR_plot_eta_095} it is clear that the secret key rates decrease by two orders of magnitude from the initial value due to the decrease in the raw rate ($R$).

We also find that in SEC, for a minimum segment length of $10$ km and a memory coherence time ($\tau_{coh}$) of 1 s, LBC yields positive secret key rates only when $\eta_m > 0.78$, whereas HBC requires $\eta_m > 0.73$. Our approach involves the full evolution of the density matrix, and performing a similar rate analysis for higher-order Binomial codes is challenging due to the exponential growth of terms in the density matrix. However, we expect a similar trend for higher orders, with a trade-off between the benefits of higher-order encoding and the coherence time of the memories.

\FloatBarrier
\section{Conclusion and outlook}

We have studied the near-term performance of a second-generation quantum repeater protocol under realistic conditions, in particular, focusing on a Binomial quantum error correction code with loss as an error model for the quantum memories at the repeater stations. In addition and more specifically, we have proposed a promising implementation using microwave cavities that achieves all four necessary key steps: state engineering, syndrome measurement, error correction, and entanglement swapping. Our simulations of a two-segment quantum repeater allowed us to examine memory loss errors and the influence of the Binomial encoding. We found that both schemes with the Lower Binomial Code (LBC) and with the Higher Binomial Code (HBC) outperform the unencoded schemes for the same coherence times of the quantum memories. The HBC, even with lower coherence times, surpassed the performance of the unencoded protocols with higher coherence times. Moreover, we observed that multiple-error correction (MEC) provides a better secret key fraction (SKF) than single-error correction (SEC). We considered two different memory interface efficiencies, $\eta_m$, and found that the SKR trends are similar for the encoded and the unencoded cases. Our methods and results offer valuable guidance for optimizing repeater protocols in upcoming experiments. Our analysis assumed ideal microwave-to-optical conversion. While such converters are not yet realized, partial implementations exist and there is also more research in this. Any non-idealities can be incorporated into the entanglement generation probability without fundamentally altering our model.

As a next step, it is interesting to analyze multi-segment repeater systems and networks. Since MEC turned out to be beneficial in the ideal case, it would be valuable to extend the analysis by including overhead losses, as there should exist an optimal number of recovery operations. This is because repeated application of error correction can itself introduce additional errors.
%Determining these optimal SEC remains an open question that requires further analysis. 
Exploring Binomial-code error correction in the context of third-generation quantum repeaters, beyond existing schemes, is a possible, promising direction as well. In this context, it would be useful to investigate state engineering in the optical domain and consider improving the $50\%$ bound for logical Bell measurements for entanglement swapping. A rigorous rate analysis in the context of $\eta_m$ would also be interesting to pursue. In summary, our findings demonstrate the promise of Binomial-code quantum error correction in advancing quantum repeater protocols and suggest important directions for future work.

\begin{acknowledgments}

S.C. thanks Simon Reiß and Evgeny Shchukin for useful discussions. 
We acknowledge support from the Deutsche Forschungsgemeinschaft (DFG, German Research Foundation)—Project-ID 429529648—TRR 306 QuCoLiMa (“Quantum Cooperativity of Light and Matter”).
S.C., F.S., and P.v.L. acknowledge funding from the Bundesministerium für Bildung und Forschung (BMBF) under the projects QR.X/QR.N, PhotonQ, and QuKuK.
S.S. and S.V.K. acknowledge funding from the Bundesministerium für Bildung und Forschung (BMBF) under the project QECHQS (Grant No. 16KIS1590K).

\end{acknowledgments}

\appendix

\section{Logical linear-optics Bell measurements}

\label{appendixa}

In this appendix, we show that a \(50\%\) success rate in entanglement swapping/Bell-state measurement (BSM) using linear optics can be achieved for lower-order Binomial codes when there is no loss on the codewords.
This means, in our repeater scheme, we have to assume that a recovery operation is performed before the entanglement swapping takes place such that all states are mapped back onto the original codespace.

The BSM requires to distinguish the logical two-qubit states given in Eq.~(\ref{BSs}).
Now first notice that $\ket{\overline{\Phi^{\pm}}}$ only contain Fock states of 0, 4, and 8 photons, 
whereas $\ket{\overline{\Psi^{\pm}}}$ only have terms with 2 and 6 photons.
We will apply a beam splitter that preserves the total photon number in each term and so
accepting only 2 or 6 photons at the detectors will unambiguously identify the two Bell states
$\ket{\overline{\Psi^{\pm}}}$. In order to discriminate among these two states, we have to examine the possible number patterns of the two-mode state at the output of the beam splitter operation, $\hat{\rm BS}$.
A 50/50 beam splitter transforms $\ket{\overline{\Psi^{\pm}}}$ as
\begin{align} 
    \ket{\overline{\Psi^{\pm}}} &= \frac{\ket{\bar{0}\bar{1}} \pm \ket{\bar{1}\bar{0}}}{\sqrt{2}} \notag \\
    &= \frac{1}{\sqrt{2}} \left( \frac{\ket{0} + \ket{4}}{\sqrt{2}} \ket{2} \pm \ket{2}
    \frac{\ket{0} + \ket{4}}{\sqrt{2}} \right) \notag \\
    &= \frac{1}{2} \left( \ket{02}\pm \ket{20} + \ket{42}\pm \ket{24} \right) \notag \\
    &\rightarrow\begin{cases}
               \frac{1}{2} \left[ \ket{20} + \ket{02} + \hat{\rm BS}\left(\ket{42} + \ket{24}\right)\right] \\
               \frac{1}{2} \,\,\left[ \sqrt{2} \ket{11} + \hat{\rm BS}\left(\ket{42} - \ket{24}\right)\right] .
            \end{cases}  
\end{align}
Here we have used $\hat{\rm BS}\ket{20} = \frac{1}{2}(\ket{20} + \ket{02}) + \frac{1}{\sqrt{2}}\ket{11}$
and $\hat{\rm BS}\ket{02} = \frac{1}{2}(\ket{20} + \ket{02}) - \frac{1}{\sqrt{2}}\ket{11}$ which means that
$\frac{1}{\sqrt{2}}(\ket{20}+\ket{02})$ remains invariant under the beam splitter transformation, while $\frac{1}{\sqrt{2}}(\ket{20}-\ket{02})$ becomes $\ket{11}$ (like an inverse Hong-Ou-Mandel effect).
Thus, click patterns `11' and `20/02' unambiguously identify the states $\ket{\overline{\Psi^{-}}}$ and
$\ket{\overline{\Psi^{+}}}$, respectively.
It then remains to be shown that also the patterns that originate from 
$\hat{\rm BS}\left(\ket{42} \pm \ket{24}\right)$ are in one-to-one correspondence to the two
different Bell states. This can easily be inferred from the following beam splitter transformations:
\begin{align}
    \hat{\rm BS}\ket{24} &=
    \frac{1}{8} \Bigl[ \sqrt{15} \ket{06} + \sqrt{10} \Ket{15} \notag \\ 
    &\quad - \Ket{24} - 2\sqrt{3} \Ket{33} 
    - \Ket{42} \notag \\
    &\quad + \sqrt{10} \Ket{51} + \sqrt{15} \Ket{60} \Bigr], \notag \\
    \hat{\rm BS}\ket{42} &=
    \frac{1}{8} \Bigl[ \sqrt{15} \ket{06} - \sqrt{10} \Ket{15} \notag \\
    &\quad - \Ket{24} + 2\sqrt{3} \Ket{33}
    - \Ket{42} \notag \\
    &\quad- \sqrt{10} \Ket{51} + \sqrt{15} \Ket{60} \Bigr].
    \label{eq:bsm2}
\end{align}
From the above equations, it is clear that the states $\hat{\rm BS}\left(\ket{42} \pm \ket{24}\right)$ produce the unique click patterns `06', `24', `42', `60' and `15', `33', `51', respectively.
As a consequence, this proves that half of the Bell states can always be unambiguously identified, thus leading to a 50\% BSM efficiency.  

\begin{figure}
    \centering
    \includegraphics[width=\columnwidth]{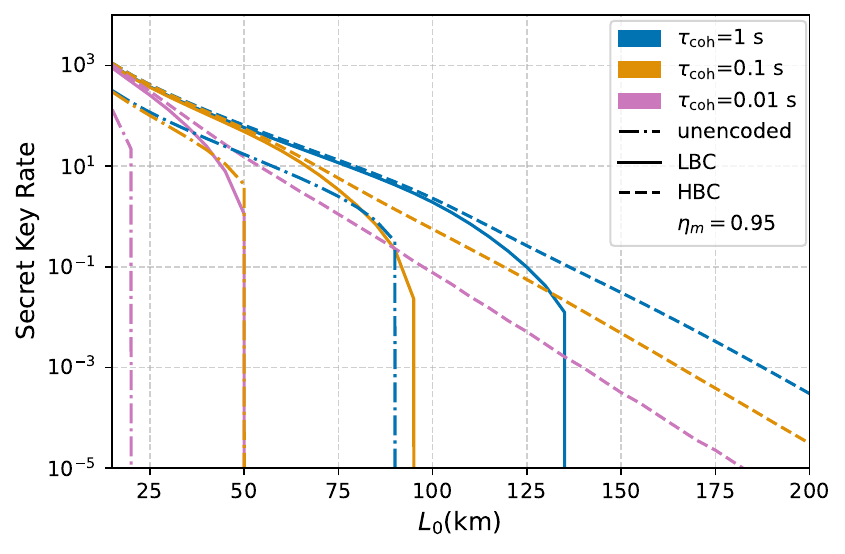}
    \caption{Comparison of secret key rates with different memory coherence times for unencoded, Lower Binomial Code (LBC) and Higher Binomial code (HBC) when SEC is performed and entanglement swapping is performed using linear optics with $50\%$ efficiency. The memory interface parameter $\eta_m=0.95$}
    \label{fig:all_optical_skr}
\end{figure}

\begin{figure}[tbh!]
    \centering
    \includegraphics[width=\columnwidth]{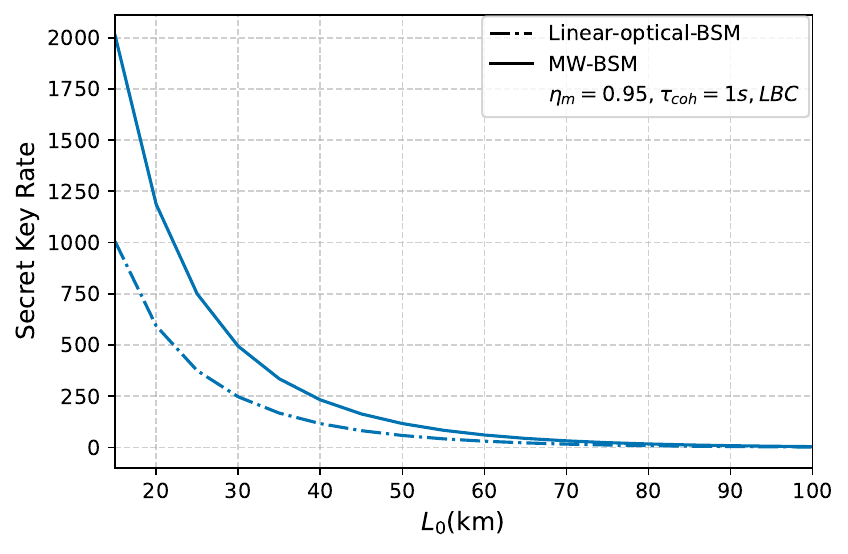}
    \caption{The comparison of SKR for all-optical and microwave cavity approach for lower order binomial code (LBC).}
    \label{fig:opt-mw}
\end{figure}

The secret key rate (SKR) for the all-optical entanglement swapping scheme is shown in Figure~\ref{fig:all_optical_skr}, while Figure~\ref{fig:opt-mw} compares the performance of the all-optical and microwave-cavity-based entanglement swapping approaches.

\section{Recovery operators}
The no-loss recovery operator for the LBC in the linear loss order, as given in the main text by Eq.~(\ref{eq:zero_recovery}),
can be equivalently written as (using the notation from Ref.~\cite{bin_code})
\begin{equation}
\begin{aligned}
       \mathscr{R}_0 &= (\sin(\log({1-\gamma}))) (\ket{E^0_{\Bar{0}}} \bra{\Bar{0}}-\ket{\Bar{0}} \bra{E^0_{\bar{0}}}) \\  
   & + (\cos(\log({1-\gamma}))) (\ket{\Bar{0}}\bra{\Bar{0}}  + \ket{E^0_{\Bar{0}}} \bra{E^0_{\Bar{0}}}) \\
   & +\ket{\Bar{1}}\bra{\Bar{1}}\,,
\end{aligned}
\end{equation}
where $\ket{E^0_{\Bar{0}}} = (\ket{0}-\ket{4})/\sqrt{2}$. Note that $\mathscr{R}_0 = \text{id}$ when $\gamma = 0$.

The no-loss recovery operator for the HBC is given by \cite{bin_code}
\begin{multline}
    \mathscr{R}_0 = \sum_{\sigma}  \sqrt{1- \frac{ |\langle \hat{B}_0  \rangle|^2 }{\beta_0}} (\Ket{W_\sigma}\bra{B^0_\sigma}-\ket{B^0_\sigma} \bra{W_\sigma}) \\
      + \frac{\langle \hat{B_0} \rangle}{\sqrt{\beta_0}} (\ket{W_\sigma}\bra{W_\sigma} + \ket{B^0_\sigma} \bra{B^0_\sigma}) \,,
\end{multline}
where $\Ket{W_\sigma}$ is the codeword, 
\begin{equation} 
    \beta_0 = \Braket{W_\sigma | \hat{B}_0^{\dagger} \hat{B}_0 | W_\sigma}, 
\end{equation} 
$\hat{B}_0 = 1 - \log(1-\gamma) \hat{n}/2 + (\log(1-\gamma))^2 \hat{n}^2/8$, and 
\begin{equation}
    \ket{B^0_\sigma}=\frac{(1-\ket{W_\sigma}\bra{W_\sigma})\hat{B}_0\Ket{W_\sigma}}{\sqrt{\beta_0-|\langle \hat{B}_0 \rangle|^2}} \,.
\end{equation}
The one-loss recovery operator for the HBC is given by 
\begin{equation}
\begin{split}
      \mathscr{R}_1 &= \ket{\Bar{0}} \bra{5} - \ket{5} \bra{\Bar{0}} \\  
   &+ \frac{1}{\sqrt{2}} \ket{\Bar{1}} \left(\bra{2}+\bra{8}\right)  -  \frac{1}{\sqrt{2}} \left(\ket{2}+\ket{8}\right) \bra{\bar{1}} \,.
\end{split}
\end{equation}
The two-loss recovery operator for the HBC is given by
\begin{equation}
\begin{split}
      \mathscr{R}_2 &= \ket{\Bar{0}} \bra{4} - \ket{4} \bra{\Bar{0}}
   + \ket{\Bar{1}}\left(\frac{1}{\sqrt{5}} \bra{1}+ \frac{2}{\sqrt{5}} \bra{7}\right)  \\ 
   &-  \left(\frac{1}{\sqrt{5}} \ket{1}+ \frac{2}{\sqrt{5}} \ket{7}\right)\bra{\bar{1}} \,.
\end{split}
\end{equation}

\section{Multiple rounds of error correction}     

\label{appendixc}

Here, we consider a scheme with multiple rounds of quantum error correction where we perform one round thereof after a fixed number of time steps $t_k$. Thus, we calculate the concatenation of the loss channel after $l$ time states followed by the recovery operation, which we will refer to as $C^l$, for $l \in \{1,\dots,t_k\}$ .
As an example let us consider that one quantum  memory needs to wait for 100 time steps until entanglement is also distributed in the other segment and we choose $t_k$ to be 30.
The resulting final noise channel is then given by $C^{10}(C^{30}(C^{30}(C^{30}(\cdot)))$. Here we can see that we have to concatenate many channels.
We do this concatenation by mapping the channel onto a vector of coefficients and the vector of coefficients after the concatenation can be obtained by multiplying the old vector of coefficients with a matrix.
The mapping works as follows: For simplicity, we consider a qubit channel, i.e., we will project $C^l$ onto a map that acts within the logical qubit space and a map that maps from the logical qubit space outside the codespace. Thus, we neglect parts mapping from outside the code space onto the codespace.
Each single-qubit channel $\mathcal{C}$ can be written in Kraus-operator representation $\sum_{j=1}^4 \sqrt{p_j}\hat{U}_j\cdot\hat{U}_j^\dagger\sqrt{p_j}$, where $p_j$ are probabilities and $\hat{U}_j$ are unitary operators.
Pauli operators form an orthogonal basis with respect to the Frobenius inner product $\langle a,\hat{B}\rangle={\rm Tr} \left(a^\dagger\hat{B}\right)$.
Thus, we can write $\hat{U}_j=\alpha_{j0}1+\alpha_{j1}\sigma_1+\alpha_{j2}\sigma_2+\alpha_{j3}\sigma_3$.
We can simplify the next expressions by defining $\sigma_0$=1 and
$\mathcal{N}=\sum_{j=0}^3\sum_{k,l}p_j\alpha_{jk}\sigma_k\cdot\sigma_l\alpha^*_{jl}=\sum_{k,l}p_{kl}\sigma_k\cdot \sigma_l$, where $p_{kl}=\sum_{j=0}^3p_j\alpha_{jk}\alpha^*_{jl}$. The coefficients $p_{kl}$ then contain the complete information of the channel. When considering the concatenation of two channels one can find that the new channel is simply given by the product of a matrix defined by the outer channel and the vector of coefficients of the inner channel. Thus, the multiple concatenations of the same channel can be represented by a matrix product.

\nocite{*}

\bibliography{apssamp}% Produces the bibliography via BibTeX.

\end{document}